# Virtual Collaborative R&D Teams in Malaysia Manufacturing SMEs

Nader Ale Ebrahim[1], Shamsuddin Ahmed, Salwa Hanim Abdul Rashid and M. A. Wazed
Department of Engineering Design and Manufacture,
Faculty of Engineering, University of Malaya
Kuala Lumpur, Malaysia
[1]e-mail: aleebrahim@siswa.um.edu.my

Zahari Taha
Faculty of Manufacturing Engineering and Management Technology,
University Malaysia Pahang,
26300 Gambang,
Pahang, Malaysia

*Abstract*—This paper presents the results of empirical research conducted during March to September 2009. The study focused on the influence of virtual research and development (R&D) teams within Malaysian manufacturing small and medium sized enterprises (SMEs). The specific objective of the study is better understanding of the application of collaborative technologies in business, to find the effective factors to assist SMEs to remain competitive in the future. The paper stresses to find an answer for a question "Is there any relationship between company size, Internet connection facility and virtuality?". The survey data shows SMEs are now technologically capable of performing the virtual collaborative team, but the infrastructure usage is less. SMEs now have the necessary technology to begin the implementation process of collaboration tools to reduce research and development (R&D) time, costs and increase productivity. So, the manager of R&D should take the potentials of virtual teams into account.

## I. INTRODUCTION

Collaboration in research and development (R&D) is becoming increasingly important in creating the knowledge that makes research and business more competitive [1]. The internet, incorporating computers and multimedia, has provided tremendous potential for remote integration and collaboration in business and manufacturing applications [2]. Web service technology also provides a unique way to application-to-application interaction over the internet [3]. Currently, many R&D organizations and teams use a specialized knowledge portal for research collaboration and knowledge management [4]. A web-based virtual collaborative team is enabling authorized users in geographically different locations to have access to the company's product data such as product drawing files stored at designated servers and carry out product design work simultaneously and collaboratively on any operating systems [5]. Despite computers' widespread use for personal applications, very few SMEs use this new phenomenon [6]. On the other hand, small and medium-sized enterprises (SMEs) which are a major part of the industrial economies [7] needs to reduce R&D time and costs in order to compete in the competitive market. Gassmann and Keupp [8] found that managers of SMEs should invest less in tangible assets, but more in those areas that will directly create their future competitive advantage (e.g., in R&D to generate knowledge, and in their employees' creativity to stimulate incremental innovations in existing technologies). One very important trend to enable new knowledge creation and transfer in and to SME's is developing virtual collaborative environments and networks to increase their innovation abilities as a single unit but also the capabilities of the network as a whole [9]. Virtuality has been presented as one solution for SMEs aiming to increase their competitiveness [10, 11]. Virtual teams reduce time-to-market [12, 13]. Lead Time or Time to market has been generally admitted to being one of the most important keys for success in manufacturing companies [14].

In line with moving trend to virtual collaborative teams in SMEs, this paper based on the survey results explore the relationship between the number of SMEs employee and the Net connection facility with virtuality. While, virtuality brings couples of advantages to SMEs, the question is raised "Why SMEs do not use virtual collaborative teams?". Based on literature and survey finding future study and suggestions are advanced.

## II. SMEs DEFINITION AND IMPORTANCE

There are many accepted definitions of SMEs, and the classifications vary from industry to industry and from country to the country [15]. Different countries adopt different criteria such as employment, sales or investment for defining small and medium enterprises [16]. The case studies employed the definition of Malaysian manufacturer sector SMEs according to Table I.

Economists believe that the wealth of nations and the growth of their economies strongly depend upon their SMEs' performance [17]. In many developed and developing countries, SMEs are the unsung heroes that bring stability to the national economy. They help buffer the shocks that come with the boom and bust of economic cycles [18, 19]. SMEs

TABLE I. DEFINITION OF MALAYSIAN MANUFACTURING SMEs [ADOPTED FROM (ALE EBRAHIM ET AL., 2009A)]

| Category of enterprise | Employee numbers | Turnover |
|---|---|---|
| Small | Between 5 to 50 employees | Between RM 250,000 (~80,000 USD) & less than RM 10 million (~3.2 million USD) |
| Medium | Between 51 to 150 employees | Between RM 10 million (~3.2 million USD)& RM 25 million (~8 million USD) |

also serve as the key engine behind equalizing income disparity among workers [20].

### III. R&D DISTRIBUTED TEAM AND SMEs

SMEs need appropriate and up-to-date knowledge in order to compete and there is a strong need to create, share and disseminate knowledge within SME's [21]. Especially, in the emerging and dynamic markets the shared knowledge creation and innovation may speed up market development [22]. The key elements in knowledge sharing are not only the hardware and software, but also the ability and willingness of team members to actively participate in the knowledge sharing processes [23]. Dickson and Hadjimanolis [24] examined innovation and networking among small manufacturing companies. They found some tentative evidence that companies operating in terms of "the local strategic network" are more innovative than those operating in terms of "the local self-sufficiency". In the beginning of R&D activities SMEs always face capital shortage and need technological assistance. Most firms today do not operate alone; they are networked vertically with many value-chain partners [25]. R&D activities are now dependent to different location drivers [26]. Most SMEs are heavily reliant on external sources, including customers and suppliers, for the generation of new knowledge [27]. SMEs of all sizes must reach out into their external environment for necessary resources [28]. In the present era of globalization, it is obvious that the survival of the SMEs will be determined first and foremost by their ability to manufacture and supply more, at competitive cost, in less delivery time, with minimum defects, using fewer resources [29]. In order to face this challenge, SMEs can reinforce knowledge to create synergies that allow firms to overcome difficulties and succeed. This may lead to new relationships between different agents to overcome scarcity and/or difficulties in gaining access to resources [30]. The combination of explosive knowledge growth and inexpensive information transfer creates a fertile soil for unlimited virtual invention [25]. Web resource services can help the enterprises to get external service resources and implement collaborative design and manufacturing [31]. Sharma and Bhagwat [29] study results reveal that IT in SMEs is still in a backseat, although the use of computers is continuously increasing in their operations.

### IV. METHOD & DATA COLLECTIONS

An online survey was conducted in the spring and summer of 2009 to identify the relationship between the number of employees, Internet connection and virtual teaming, among the Malaysian manufacturing SMEs. The on line questionnaire was distributed through the e-mail to Malaysian manufacturing SMEs. Two thousand and sixty eight email addresses collected from Malaysian SME Business Directory [32] and questionnaires were sent to manufacturing SMEs. The online system received replies from 356 entities were received the email and clicked the link, within the desired timeframe. Participants were directed to a website, and the survey was completed on-line. The rapid expansion of Internet users has given web-based surveys the potential to become a powerful tool in survey research [33]. Denscombe [34] findings encourage social researchers to use web-based questionnaires with confidence and the data produced by web-based questionnaires is equivalent to that produced by paper-based questionnaires. Other authors stressed the data provided by Internet methods are of at least as good quality as those provided by traditional paper-and-pencil methods [35, 36]. The survey was first tested with 12 expert people, then adjusted and distributed.

TABLE II. SUMMARIZED ONLINE SURVEY DATA COLLECTION

| | |
|---|---|
| Numbers of emails sent to Malaysian Firms | 2068 |
| Total Responses (Click the online web page) | 356 |
| Total Responses / Received questionnaire (%) | 17.2 |
| Total Completed | 74 |
| Total Completed / Received questionnaire (%) | 20.8 |

Finally, a questionnaire was distributed to 356 Malaysian manufacturing SMEs. The main target group regards the organization's size and field of industry was, managing director, R&D manager, new product development manager, project and design manager and right people who were most familiar with the R&D issue in the organizations. 74 usable questionnaires were received, representing a 20.8 percent return rate. The response rate was satisfactory since accessing the managers is usually difficult. Table II summarized online survey data collection. 42 SMEs were met the criteria of this research so the rest of responded took away from analysis. Descriptive statistics were used to analyze the responses. Table III shows the frequency of using virtual teams among the sample Malaysian SMEs.

### V. SURVEY RESULTS AND DISCUSSION

From the data presented in Table III, we see that although, virtual teams' application in manufacturing SMEs is still in infancy but virtual teaming is becoming accepted in Malaysian manufacturing SMEs. One out of three companies uses virtual teams. A cross-tabulation descriptive statistics employed to find the frequency and relationship between the virtuality, number of employees and the type of internet connections, as illustrate in Table IV. The result shows that small SMEs employed virtual collaborative teams but medium sized SMEs in the sample did not use virtual teams, although they have sufficient internet connection facilities.

TABLE III. CROSS-TABULATION BETWEEN COUNTRY AND VIRTUAL TEAM

| | Using Virtual Team | | Total |
|---|---|---|---|
| | *Yes* | *NO* | |
| **Count** | 14 | 28 | 42 |
| **%** | 33.3% | 66.7% | 100.0% |

TABLE IV.  CROSS-TABULATION BETWEEN VIRTUAL TEAM, NUMBER OF EMPLOYEES AND THE TYPE OF INTERNET CONNECTIONS

| With Virtual Teams | Internet connection | Number of Employees Count (%) | | | | | | Total |
|---|---|---|---|---|---|---|---|---|
| | | 10< | 11-20 | 21-30 | 31-50 | 51-100 | 101-150 | |
| Yes | *Broadband Network* | 7 | 3 | 1 | | | | 11 |
| | *DSL (Digital Subscriber Line)* | 2 | 0 | 0 | | | | 2 |
| | *Direct Satellite Connection* | 0 | 1 | 0 | | | | 1 |
| Total | | 9 (64.3) | 4 (28.6) | 1 (7.1) | | | | 14 (100) |
| No | *Do not have internet connection* | 0 | 0 | 0 | 1 | 0 | 0 | 1 |
| | *Dial Up* | 1 | 1 | 0 | 0 | 0 | 0 | 2 |
| | *ISDN* | 0 | 1 | 0 | 0 | 0 | 0 | 1 |
| | *Broadband Network* | 8 | 6 | 1 | 3 | 2 | 1 | 21 |
| | *DSL (Digital Subscriber Line)* | 0 | 1 | 0 | 0 | 1 | 0 | 2 |
| | *Direct Satellite Connection* | 1 | 0 | 0 | 0 | 0 | 0 | 1 |
| Total | | 10 (35.7) | 9 (32.1) | 1 (3.6) | 4 (14.3) | 3 (10.7) | 1 (3.6) | 28 (100) |

## A. Correlation Analysis

Due to the lack of normality of collected data the Spearman non-parametric statistical correlations for ordinal data were, employed. Table V shows, the significant correlation coefficient between virtuality and number of employees (p = 0.035 and r = 0.327). The result shows that virtuality and number of employees (0.327) has the strongest relationship among virtuality, Internet connections and number of employees. There is not significant correlation between Internet connections, number of employees and virtuality.

The research findings indicate that both SMEs with employing virtual teams and not, equally has access to the internet connection's facilities. No correlation was found between the number of Employees in SMEs and Internet connections in the Malaysian manufacturing sector. Spearman's rho correlation coefficient was -0.090.

TABLE V.  CORRELATIONS BETWEEN VIRTUALITY, INTERNET CONNECTIONS AND NUMBER OF EMPLOYEES

| | | No. of Employees | Virtuality | Internet connections |
|---|---|---|---|---|
| **No. of Employees** | *Correlation Coefficient* | 1.000 | 0.327* | -0.090 |
| | *Sig. (2-tailed)* | . | 0.035 | 0.571 |
| | *N* | 42 | 42 | 42 |
| **Virtuality** | *Correlation Coefficient* | 0.327* | 1.000 | -0.240 |
| | *Sig. (2-tailed)* | 0.035 | . | 0.126 |
| | *N* | 42 | 42 | 42 |

*. CORRELATION IS SIGNIFICANT AT THE 0.05 LEVEL (2-TAILED).

## VI. CONCLUSION

The research and development requires higher levels of expertise within SMEs. Exchange knowledge and expertise can be created across virtual R&D teams. In principle, virtual teams could allow rapid decision-making to operate within SMES, regardless of the geographical location of its members. Although the infrastructure is ready for almost all SMEs, only one third of SMEs use internet connection facilities for establishing virtual R&D teams. So, the manager of R&D should take the potentials of virtual teams into account. Data from the Malaysian manufacturing SMEs sources shows, technologically SMEs capable of performing the virtual collaborative team. Despite the enormous benefaction of employ virtual R&D teams in manufacturing SMEs, applying the virtual teams by most enterprises, is still at its infancy.

This study is probably the first to present an empirical study on virtual R&D teams, which was limited to Malaysian manufacturing SMEs. The future research needs to investigate the correlations between the number of employees, virtuality and Internet connections by a larger sample from different sectors. The theme of virtual collaborative R&D teams has not been much explored and researchers in this field are encouraging to do more studies.

## REFERENCES

[1] M. A. Shafia, et al., "Innovation Process is Facilitated in Virtual Environment of R&D Teams," in International Conference on Education and New Learning Technologies (EDULEARN09), Barcelona, Spain, 2009, pp. 2157-2166.

[2] H. Lan, et al., "A web-based manufacturing service system for rapid product development," Computers in Industry, vol. 54, pp. 51 - 67  2004.

[3] M. Xiaozhen, et al., "Collaborative Product Development in SMEs: A case study for CRC," in IEEE International Conference on Industrial Informatics(INDIN'06), Singapore, 2006, pp. 938-942.

[4] H. J. Lee, et al., "A contingent approach on knowledge portal design for R&D teams: Relative importance of knowledge portal functionalities," Expert Systems with Applications, vol. ARTICLE IN PRESS, 2008.

[5] H. F. Zhan, et al., "A web-based collaborative product design platform for dispersed network manufacturing," Journal of Materials Processing Technology, vol. 138, pp. 600-604, 2003.

[6] N. Ale Ebrahim, et al., "Dealing with Virtual R&D Teams in New Product Development," in The 9th Asia Pacific Industrial Engineering & Management Systems Conference and the 11th Asia Pacific Regional Meeting of the International Foundation for Production Research, Nusa Dua, Bali - Indonesia, 2008, pp. 795-806.

[7] T. R. Eikebrokk and D. H. Olsen, "An empirical investigation of competency factors affecting e-business success in European SMEs," Information & Management, vol. 44, pp. 364-383 2007.

[8] O. Gassmann and M. M. Keupp, "The competitive advantage of early and rapidly internationalising SMEs in the biotechnology industry: A knowledge-based view," Journal of World Business, vol. 42, pp. 350-366, 2007.

[9] M. Flores, "IFIP International Federation for Information Processing," in Network-Centric Collaboration and Supporting Fireworks. vol. 224, Boston: Springer, 2006, pp. 55-66.

[10] T. Pihkala, et al., "Virtual organization and the SMEs: a review and model development," Entrepreneurship & Regional Development, vol. 11, pp. 335 - 349, 1999.

[11] N. A. Ebrahim, et al., "Critical factors for new product developments in SMEs virtual team," African Journal of Business Management, vol. 4, pp. 2247-2257, Sep 2010.

[12] A. May and C. Carter, "A case study of virtual team working in the European automotive industry," International Journal of Industrial Ergonomics, vol. 27, pp. 171-186, 2001.

[13] N. A. Ebrahim, et al., "SMEs; Virtual research and development (R&D) teams and new product development: A literature review," International Journal of the Physical Sciences, vol. 5, pp. 916-930, Jul 2010.

[14] M. Sorli, et al., "Managing product/process knowledge in the concurrent/simultaneous enterprise environment," Robotics and Computer-Integrated Manufacturing, vol. 22, pp. 399–408, 2006.

[15] N. O'Regan and A. Ghobadian, "Testing the homogeneity of SMEs - The impact of size on managerial and organisational processes," European Business Review, vol. 16, pp. 64-79, 2004.

[16] M. Ayyagari, et al., "Small and Medium Enterprises Across the Globe," Small Business Economics, vol. 29, pp. 415-434, 2007.

[17] H. H. Schröder, "Past, Present and Future of Knowledge Integration," in Knowledge Integration-The Practice of Knowledge Management in Small and Medium Enterprises, A. Jetter, et al., Eds.: Physica-Verlag HD, 2006, pp. 175-191.

[18] N. A. Ebrahim, et al., "Virtual R&D teams and SMEs growth: A comparative study between Iranian and Malaysian SMEs," African Journal of Business Management, vol. 4, pp. 2368-2379, Sep 2010.

[19] N. Ale Ebrahim, et al., "Virtual R & D teams in small and medium enterprises: A literature review," Scientific Research and Essay, vol. 4, pp. 1575–1590, December 2009.

[20] T. Y. Choi, "Korea's Small and Medium-Sized Enterprises: Unsung Heroes or Economic Laggards?," Academy of Management Executive, vol. 17, pp. 128-129, 2003.

[21] M. B. Nunes, et al., "Knowledge management issues in knowledge-intensive SMEs," Journal of Documentation, vol. 62, pp. 101-119, 2006.

[22] K. Blomqvist, et al., "Towards networked R&D management: the R&D approach of Sonera Corporation as an example," R&D Management, vol. 34, pp. 591-603, 2004.

[23] B. Rosen, et al., "Overcoming Barriers to Knowledge Sharing in Virtual Teams," Organizational Dynamics, vol. 36, pp. 259–273, 2007.

[24] K. E. Dickson and A. Hadjimanolis, "Innovation and networking amongst small manufacturing firms in Cyprus," International Journal of Entrepreneurial Behavior & Research, vol. 4, pp. 5-17, 1998.

[25] R. E. Miles, et al., "TheFuture.org " Long Range Planning, vol. 33, pp. 300-321, 2000.

[26] M. Von Zedtwitz and O. Gassmann, "Market versus technology drive in R&D internationalization: four different patterns of managing research and development," Research Policy, vol. 31, pp. 569-588, 2002.

[27] O. Jones and A. Macpherson, "Inter-Organizational Learning and Strategic Renewal in SMEs," Long Range Planning, vol. 39, pp. 155-175, 2006.

[28] P. H. Dickson, et al., "Opportunism in the R&D alliances of SMES: The roles of the institutional environment and SME size," Journal of Business Venturing, vol. 21, pp. 487–513 2006.

[29] M. K. Sharma and R. Bhagwat, "Practice of information systems: Evidence from select Indian SMEs," Journal of Manufacturing Technology Management, vol. 17, pp. 199 - 223, 2006.

[30] J. O. Gomez and M. Simpson, "Achieving competitive advantage in the Mexican footwear industry," Benchmarking: An International Journal, vol. 14, pp. 289-305, 2007.

[31] B. Dong and S. Liu, "Implementation of Web Resource Service to Product Design " in International Federation for Information Processing -Knowledge Enterprise: Intelligent Strategies in Product Design, Manufacturing, and Management. vol. 207, K. Wang, et al., Eds., Boston: Springer 2006.

[32] SMIDEC. (2008). Malaysian SME Business Directory (4th ed.). Available: http://www.yellowpages.com.sg/newiyp/jsp/osme/OsmeSearchBox.jsp

[33] S. J. Sills and C. Song, "Innovations in Survey Research: An Application of Web-Based Surveys," Social Science Computer Review, vol. 20, pp. 22-30, February 2002.

[34] M. Denscombe, "Web-Based Questionnaires and the Mode Effect: An Evaluation Based on Completion Rates and Data Contents of Near-Identical Questionnaires Delivered in Different Modes," Social Science Computer Review, vol. 24, pp. 246-254, May 2006.

[35] S. D. Gosling, et al., "Should We Trust Web-Based Studies? A Comparative Analysis of Six Preconceptions About Internet Questionnaires," American Psychologist, vol. 59, pp. 93-104, 2004.

[36] E. Deutskens, et al., "An assessment of equivalence between online and mail surveys in service research," Journal of Service Research, vol. 8, pp. 346-355, May 2006.